# Integrating Existing Software Toolkits into VO System

Chenzhou Cui[*a], Yongheng Zhao [a], Xiaoqian Wang [a], Jian Sang [a], Ze Luo [b]

[a]National Astronomical Observatories, CAS,20A Datun Road, Chaoyang District, Beijing 100012, China
[b]Computer Network Information Center, CAS, 4 Nansijie, Zhongguancun, Haidian District, Beijing 100080, China

**ABSTRACT**

Virtual Observatory (VO) is a collection of interoperating data archives and software tools. Taking advantages of the latest information technologies, it aims to provide a data-intensively online research environment for astronomers all around the world.

A large number of high-qualified astronomical software packages and libraries are powerful and easy of use, and have been widely used by astronomers for many years. Integrating those toolkits into the VO system is a necessary and important task for the VO developers.

VO architecture greatly depends on Grid and Web services, consequently the general VO integration route is "Java Ready – Grid Ready – VO Ready". In the paper, we discuss the importance of VO integration for existing toolkits and discuss the possible solutions. We introduce two efforts in the field from China-VO project, "gImageMagick" and " Galactic abundance gradients statistical research under grid environment". We also discuss what additional work should be done to convert Grid service to VO service.

## 1. INTRODUCTION

Virtual Observatory (VO) is the result of combination of Astronomy and the latest information technologies. VO is driven by astronomical research and education, but is enabled by the latest technical progresses, especially those in computer science and network.

One of the final goals of different VO projects from different countries and regions is to setup an International Virtual Observatory (IVO). While, the basic aim of the IVO is to provide uniform resource access to its users. The resources provided by the IVO are all kinds of resources related to astronomical research, education and outreach, including but not limited to observational and simulating data, astronomical literatures, software tools, data storage space, computing power, network bandwidth, and even astronomical telescopes and other instruments.

At the early stage of VO research and development, interconnection and interoperation among different datasets and different VO projects are primary goals. Every year, International Virtual Observatory Alliance (IVOA) will arrange two specific meetings on interoperation.

Although all kinds of astronomical data, for example sky survey data productions, personal observations and simulation results, are important resources for the IVO, software tools to process and analyze these data are also key components, or even more important parts. Extensively applications are key criteria to judge whether the IVO is successful.

---

[*] ccz@bao.ac.cn; phone 86 10 64841695; fax 86 10 64878240; www.lamost.org/~cb

The IVO will provide astronomers all around the world a brand new online research environment. During its R&D, many new programs, libraries and toolkits will be developed to meet the requirements for huge data processing, analyzing, and visualizing. However, software has been used in Astronomy for more than 50 years[1]. A lot of excellent software have been developed and widely used. For example, comprehensive packages for data processing, like IRAF, MIDAS and AIPS++; specific toolkits for FITS operations, like fv, FTOOLS, and FITSIO/CFITSIO; astronomical image viewers, like DS9, SAOImage; very useful libraries like PGPLOT, IDL Astronomy Users Library and Numerical Recipes. Telescope Data Center[2] at Smithsonian Astrophysical Observatory (SAO) and ASDS (Astronomical Software & Documentation Service) [3] at Space Telescope Science Institute (STScI) are maintaining very nice on-line astronomical software resource lists. These popular packages are frequently used by astronomers in their daily work. It is impossible to imagine `doing astronomy' without them. Furthermore, they are also very precious legacy for future astronomers and us. It costs astronomers and other developers tens of years to develop and maintain those packages.

As a basic infrastructure for the future astronomy research, it is necessary for the IVO to provide similar or enhanced functions as today's software can. It will be quite laborious work to realize functions of existing software in VO environments from scratch. Fortunately, VO is also an evolution of existing technologies and resources. It can take advantages of those existing packages and integrating them into itself.

Software integration strongly depends on the architecture and implement of the VO infrastructure. To design a uniform architecture for the whole IVOA projects is very difficult or impossible, although the IVOA is trying to do so [4]. Different projects will prefer different architectures, and especially implements. Varies of programming languages, technologies are adopted.

Chinese Virtual Observatory (China-VO) is a consortium initiated by National Astronomical Observatory of China. Its mission is to complete the part of the IVO in China, and provide its user uniform access to national and international astronomical resources. At present, the China-VO is still in its early research phase. It will adopt Grid, especially Globus Toolkit (GT) as the infrastructure of its platform. Integration, in the following sections of the paper, mainly means how to integrate existing software into the Grid based China-VO testbed.

The paper is organized as follows. In Section 2, a brief overview is given of general approaches for legacy application integration. Section 3 describes our experiences and lessons on ImageMagick VO integration. How we convert from a traditional Galactic evolution statistic research application into a set of Grid services is introduced in Section 4. We conclude in Section 5 with further discussion about some other related problems on integration.

## 2. GENERAL INTEGRATING APPROACHES

To integrate existing software into Grid based VO systems, the step before VO integration is converting those software to be Grid-enabled, i.e. Grid integration. Java's object-oriented features, platform independence, and numerous APIs for tasks such as network programming, XML processing, make it a powerful and increasingly popular language for developing Grid-based scientific applications. Interface scheme provides an elegant way to separate design and implementation[5], thus modularization and reuse will be feasible. The integration of legacy codes and applications into Java based interfaces is often required by Grid application developing.

If the existing code is written in Java, Grid integration is easy to perform, as it requires writing some new Java classes to provide additional grid methods for job creation, submission, monitoring and so on. However, C and FORTRAN are the programming languages used extensively, and most of existing software and libraries are written in C and FORTRAN. Consequently, the step before Grid integration is making them for "Java-ready", which means to incorporate software and applications in C and FORTRAN with Java.

Generally, the goal of VO integration can be reached by 3 steps: Java-ready, Grid-ready and VO-ready. An ideal VO-integrated software should be able to work as and interact with native VO services. However, what is a native VO service is still unclear. There isn't an official definition for VO service from the IVOA. In the following sections, we mainly focus our topics on the first two steps. In the last part of the paper, we will give further discussion on VO integration.

Integration efforts can be separated into two classes: specific tactics and problem-solving environments (PSE). The first class is to integrate existing software one by one, usually by hand, which is suitable for a small group to explore technical feasibility. While, the main goal of the later class is to develop a comprehensive environment to provide a fully or semi automatic solution for the wrapping of legacy codes. This approach is "once and forever", and will be more efficient and economic in the long run. But the original developing work is heavy. Furthermore, no one can ensure it is the best solution for all the software. Several C to Java or even to Grid environments have been presented, like JACAW&MEDLI[6], GAT[7], GAF4J[8], and JCI[9]. A working group in the AstroGrid project is also developing Common Execution Architecture (CEA) [10], which is an environment to send messages to and from applications so that these applications can run in Virtual Observatory.

Two main approaches have been adopted in applying Java to legacy codes: rewriting and wrapping. In the first approach, packages previously written in C or FORTRAN are completely rewritten in Java. In the second approach, legacy packages are retained and Java Native Interface (JNI) or simple local program call is used to integrate native methods into Java. This may not always be an optimal or elegant solution, but it is necessary when large libraries are not immediately available in Java and probably the only way for large binary packages without source codes.

Besides the two mainstream approaches, "automatically translating C/FORTRAN codes into JAVA" is still under research by some developers and companies. Some tools, like f2j (FORTRAN-to-Java Converter) [11], c2j (C to Java Converter) [12], Jazillian (a tool to translate C source code to Java source code) [13], Alma (Translate anything to Java) [14], are available, but code converting problem is too complex and difficult to be solved perfectly. We will not give further discussion on this approach in the paper.

In addition to the main C-Java-Grid-VO route, there are some offshoots for integration, for example "software bus" and "C# — .Net".

Define a "software bus" where both legacy modules and grid-enabled modules can be plugged easily. One example of this effort comes from European CrossGrid project, i.e. Grid Visualization Kernel (GVK) [15]. GVK is a middleware, which aims to enable the use of visualization services within computational grids. The idea of GVK is to encapsulate modules of an arbitrary visualization pipeline and distribute these modules to a number of computing elements across the grid. Python is selected for their software bus developing. Although Python is a scripting language; it could be seen as framework where you can plug easily "extension modules" in binary form implemented using other languages. The API for Python extension modules is quite simple and flexible. The GT3.x also provides interfaces for Python. "Software Bus" approach is a good compromise for complex grid applications and fast prototype developing. However, from the long run, secondary converting work for Grid will be needed. Furthermore, mixing several kinds of programs and libraries together may pose serious technical problems.

Just as the undetermined condition of the VO service, Grid is undergoing quick evolution. Java is not the only programming language and Globus Toolkit is not the standard implementation of OGSA and WSRF. Using Microsoft

C#, Vivek Haridas, etc. developed a wrapper over one of the very popular FITS libraries, CFITSIO[16] and integrated the library into Microsoft .Net platform.

## 3. GIMAGEMAGICK, GRID-ENABLED IMAGEMAGICK

### 3.1. ImageMagick, good candidate for integration

ImageMagick [17] is an open-source but robust collection of tools and libraries to read, write, and manipulate an image in many image formats, including not only most of the popular formats like TIFF, JPEG, PNG, and GIF, but also some astronomer interested formats like PS, EPS, GPLT, HTML, PDF, and especially FITS. ImageMagick also supports a number of image format specifications, which refer to images prepared via an algorithm, or input/output targets, like GRADIENT, and HISTOGRAM.

With ImageMagick, you can convert the format of an image to another; resize, rotate, sharpen, color reduce, or add special effects to an image; create montage or composite for several separate images. You can also create images dynamically. Those features are very useful for VO visualization implementation. For example, search results from VO image access services are generally files in FITS, which cannot be displayed on webpage directly (, before FITS is accepted as one of MIME media types[18] and obtains wide support from web browser distributions). Using ImageMagick, these results can be converted into web browser supported formats, like JPG, GIF or PNG, on-the-fly or extracted into thumbnails for quickly browsing.

In addition to those powerful functions, ImageMagick is a good candidate for technical exploration in Grid integration. First, image-processing operations are available from command lines, which is easy for wrapping. Second, friendly APIs are available in many programming languages, including C, C++, Perl, Java, PHP, Python, Ruby, etc., which provides a very nice platform to compare different integration methods.

### 3.2. Wrapping

Recently, as an effort in VO integration, we migrated the ImageMagick to GT 3.2 environment. "Wrapping" approach is used here to encapsulate the ImageMagick as a grid service, i.e. gImageMagick, and provide online functions. The whole procedure is implemented as four parts: Browser Client, gImageMagick Client, gImageMagick Server, and ImageMagick core. The workflow structure of the gImageMagick is shown in Fig. 1.

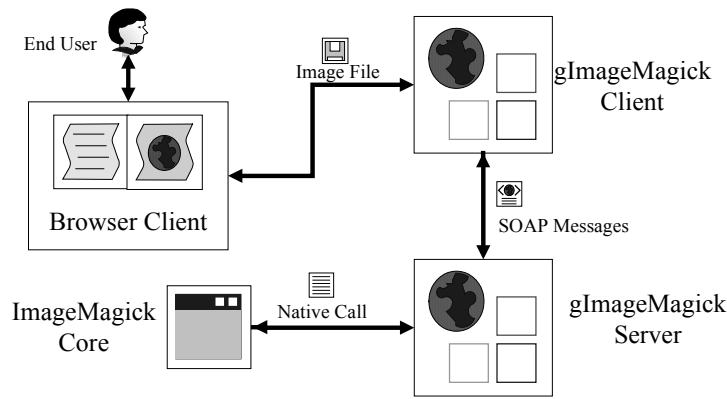

Fig.1. gImageMagick workflow

1) Browser Client is developed using JSP and Java Servlet. It has two main interfaces, one for receiving end user's input parameters, i.e. source image location, destination format, and other options; the other for retrieving converted results from the gImageMagick Client and displaying them to the end user.

2) gImageMagick Client is a grid service. It has also two main functions, to invoke the gImageMagick Server service according to input requires received from the Browser Client, and to return converted results received from the gImageMagick Server to the Browser Client.

3) gImageMagick Server is the key role for enabling the ImageMagick as a Grid service, which provides standard grid interfaces through wrapping the original ImageMagick package. Three functions are provided by the component, accepting invoking require from the gImageMagick Client, calling native ImageMagick execution program, returning results to the gImageMagick Client.

4) ImageMagick core is the original package from ImageMagick project, which is either an execution command or a shared library. In our present work, only execution command is used.

### 3.3. VO extension

For a Grid Application Developer, a software author who develops grid application by taking advantages of the available grid services, only the gImageMagick Server is needed. At present, the connection between the gImageMagick Client service and Server service is setup by direct URL endpoint. In the next step of our work, registry interface will be implemented, so that one can discover the service from a Grid Registry server, such as China-VO resource center. Furthermore, we plan to integrate the gImageMagick service with China-VO data access service[19], so that retrieved FITS images can be converted into JPG or GIF images on-the-fly and displayed to users. In our future work, we will also try to use share library and java package to compare different methods and performance after integration.

## 4. GALACTIC ABUNDANCE GRADIENTS STATISTICAL RESEARCH UNDER GRID ENVIRONMENT

### 4.1. Original implementation

This demonstration is based on our previous work[20]. Chemical evolution of the Galaxy is one of the important research fields in astronomy. Through direct observation and simple data processing, dynamic data, i.e. position and space velocity, and abundances data for a star can be obtained. Galactic mass distribution models provided by astronomers can be adopted as potential function. Through numerical integration, taking position and space velocity data as input, stellar orbits in the Galaxy and orbital parameters can be calculated. Then we can study abundance distribution and evolution in the Galaxy by federating stellar abundances with orbital parameters.

When we did the research three years ago, position and space velocity data for program input were prepared by hand. Orbital calculating program was written in C programming language by Chenzhou Cui, etc. Data analyzing and visualization work were done with commercial products, Origin Pro.

### 4.2. Service oriented implementation

The above work way is a personal research style, and hard for others to utilize. In order to do this research work under Grid environments, and then run the procedure on China-VO platform, we re-wrote required programs in Java and deployed them as Grid services. "Rewriting" approach mentioned in Section 2 was adopted. We changed the structure of the procedure from procedure-oriented to service-oriented to provide standard Grid service interfaces. Topology of this demonstration is shown in Fig. 2.

Currently, implemented Grid services include data source service, stellar orbit calculating service and statistical analysis service. Those Grid services constituted a Grid application system. Data source service is a data access service, which supports unified access to distributed and heterogeneous databases. Stellar orbit calculating service is a computing service, which calculates stellar orbits in the Galaxy. Statistical analysis service includes several sets of build-in algorithms to carry out correlation analysis. During the present prototype, service registry and discovery mechanism and workflow functions are not included. The whole procedure is guided by user's mouse clicking on his web browser. Four steps are required as follows.

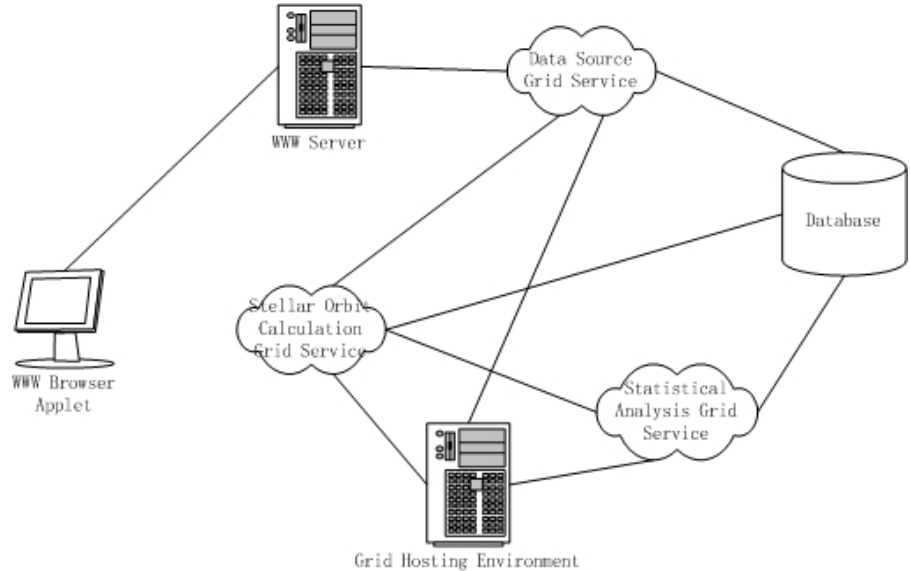

Fig.2. Topology of Galactic abundance gradients statistical research

1) Input data upload. User uploads two files in CSV format to provide stellar abundance data and dynamic data. Data Source Grid Service receives the uploaded files and stages them into a temporal database.
2) After source data have been ready in database, Stellar Orbit Calculation Grid Service will be invoked to calculate orbits for input samples using uploaded dynamic data and pre-defined Galactic mass distribution function.
3) Then, it is time to invoke the Statistical Analysis Grid Service to carry out correlation analyzing by federating orbit parameters derived from the above step with uploaded stellar abundances. Before the analysis operation, a simple cross-math operation will be run to match abundance data of a stellar with its orbital parameters.
4) The result of statistical analysis is shown in a Java Applet on user's browser. Fig. 3 is a snapshot of the result.

**4.3. Lessons learned**

The main task of converting the above demonstration from traditional procedure to grid-enabled one is program re-writing. Rewriting with Java will get OS independency, Grid compliancy in the maximum degree. It will pave the way for future extension, for example VO integration.

The orbital calculation program in the demonstration is a small program with less than 1000 line codes, which is not very difficult to rewrite. However, if the program for rewriting is large packages, for example MIDAS, the situation will be completely different.

## 5. FURTHER DISCUSSION ABOUT VO INTEGRATION

Integrating existing software and applications into VO systems is an important task for the VO developers. Whether these popular toolkits can be integrated, determines the acceptance degree of the VO in a large scope. In above sections, we described the necessity and the importance of existing software for the IVO, discussed several potential integrating approaches and introduced two of our efforts in the field, "gImageMagick" and "Grid enabled Galactic abundance gradients statistical research".

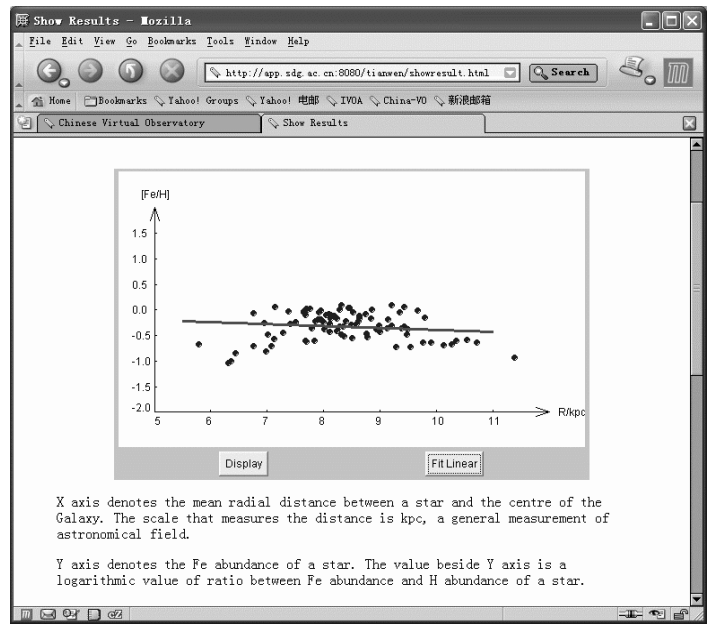

Fig. 3. Snapshot of the graphic interface displaying statistical result

Here, I must call your attention that in the above sections our discussions about integration are mainly focused on Grid integration, which is not the same as the VO integration. From the view of the China-VO, VO is an advanced application running under Grid environments. In order to integrate legacy codes into VO environments and interoperate with other VO native services, some additional work must be done. These additional work are mainly about implementation of the IVOA standards and specifications.

In order to ensure the interoperability of the different Virtual Observatory projects currently underway, many research areas have been defined in the IVOA community, like Registry, Data Access Layer, Data Model, Content Description, Query Lauguage, VOTable, and Grid & Web Services. Some specifications and recommendations have been defined or are under definition and research by corresponding working groups. Among these specifications, some are directly related with VO integration, for example " Standard Interfaces of IVOA Grid & Web Services", IVOA Data Model, Data Exchange Format ("VOTable"), Resource & Service Metadata. Different applications will need different specifications. For example, when you integrate GetImage[21], the binary search engine provided by the STScI for DSS image archive, supports for VOQL and SIAP are required.

Another important issue for VO integration is the role of common application environments as mentioned in Section 2. We don't use any environment in our efforts introduced in the above sections. Adopting developing environments will improve working efficiency, but it will bring other problems. For example, new dependency on the adopted environments. How long can one environment survive for use? How about the interoperability and compliance of these environments? Furthermore, developing environments can't do everything for you. In our opinion, a compromissary solution is to use wrapping tools for some standard work and to finish remaining specific work by hand. Manual work is always necessary.

Even though all of the above technical obstacles have been solved, it still doesn't mean that all of the existing software can be integrated into VO environments and run as VO services. Both Grid and VO are service-oriented architecture (SOA). This intrinsic character of the VO determines that not all the applications are good candidates for VO services.

Generally, small and encapsulated software with simple IO interfaces can be easily atomized and are possible candidates for VO services. However, large and complex packages with heavily "GUI" based user interfacing will be very hard to integrate. Discussion about which kind of applications are good candidates for the VO is out of the scope of the paper. A list of criterions for good Grid applications is given in the IBM Grid Redbook[22], which is also a useful reference to select candidates for VO applications.